%% file: main.tex
\documentclass[letterpaper,twocolumn,10pt]{article}
\usepackage{usenix}
\usepackage[english]{babel}
\usepackage{subcaption}
\usepackage{amsmath}
\usepackage{multirow}
\usepackage{tikz}
\usetikzlibrary{arrows}

\usepackage[subtle, lists=tight]{savetrees}
\usepackage[small, compact]{titlesec}
\usepackage{titling}
\titlespacing*{\section}{0pt}{4pt}{3pt}
\titlespacing*{\subsection}{0pt}{3pt}{3pt}
\titlespacing*{\subsubsection}{0pt}{3pt}{3pt}

\usepackage{xcolor,colortbl}

\usepackage{xspace}

\newcommand{\name}[0]{Strokkur\xspace} %

\begin{document}
\date{\vspace{-\baselineskip}}

\title{\Large \bf Coded Transaction Broadcasting for High-throughput Blockchains}

\author{
{\rm Lei Yang}\\
MIT CSAIL
\and
{\rm Yossi Gilad}\\
Hebrew University of Jerusalem
\and
{\rm Mohammad Alizadeh}\\
MIT CSAIL
} %

\maketitle

\begin{abstract}
High-throughput blockchains require efficient transaction broadcast mechanisms that can deliver transactions to most network nodes with low bandwidth overhead and latency. Existing schemes coordinate transmissions across peers to avoid sending redundant data, but they either incur a high latency or are not robust against adversarial network nodes. We present \name{}, a new transaction broadcasting mechanism that provides both low bandwidth overhead and low latency. The core idea behind \name{} is to avoid explicit coordination through randomized transaction coding. Rather than forward individual transactions. \name{} nodes send out codewords---XOR sums of multiple transactions selected at random. Since almost every codeword is useful for the receiver to decode new transactions, \name{} nodes do not require coordination, for example, to determine which transactions the receiver is missing. \name{}'s coding strategy builds on LT codes, a popular class of rateless erasure codes, and extends them to support multiple uncoordinated senders with partially-overlapping continual streams of transaction data. \name{} introduces mechanisms to cope with adversarial senders that may send corrupt codewords, and a simple rate control algorithm that enables each node to independently determine an appropriate sending rate of codewords for each peer. Our implementation of \name{} in Golang supports 647k transactions per second using only one CPU core. Our evaluation across a 19-node Internet deployment and large-scale simulation show that \name{} consumes $2$--$7.6\times$ less bandwidth than the existing scheme in Bitcoin, and 9$\times$ lower latency that Shrec when only 4\% of nodes are adversarial. 
\end{abstract}
\input{introduction}
\input{related-works}

\input{overview}
\input{design2}

\input{evaluation}
\input{conclusion}

\bibliographystyle{plain}
\bibliography{ref}

\end{document}

%% file: introduction.tex
\section{Introduction}
\label{sec:intro}

Blockchains have risen in popularity in recent years with the emergence of decentralized applications such as cryptocurrencies, smart contracts, non-fungible tokens (NFTs), decentralized finance, and more. At its core, a blockchain network consists of thousands of geodistributed \textit{nodes} that collectively implement a replicated state machine. The nodes establish a peer-to-peer network and participate in a consensus algorithm to agree on the ordering of \textit{transactions} (e.g., money transfers) that apply updates to the state machine. To perform a transaction, a client sends it to one or more nodes in a blockchain network, which then {\em broadcast} the transaction throughout the network. Most blockchain consensus protocols periodically selects a node to propose an ordered batch of transactions (called a {\em block}) to be executed next. Block proposers can be chosen in many ways (e.g., via proof-of-work~\cite{bitcoin}, verifiable random functions~\cite{algebraic}, etc.), but crucially, the identity of the proposer is not known beforehand to prevent targeted attacks. Therefore, it is important to broadcast new transactions to reach most network nodes with low latency so they can be executed in a timely manner. 

Most existing blockchain networks rely on simple transaction broadcasting mechanisms. For example, an Ethereum node simply floods new transactions to all peers~\cite{ethereum-measure}. Although obviously inefficient\,---\,flooding delivers a transaction up to 50 times per node since a node by default connects to 50 peers in the Ethereum network~\cite{ethereum-measure}\,---\,the throughput of most blockchains is quite low. The Ethereum network, for example, has a throughput of 20 transactions per second (tps), or tens of Kbps, so even a 50$\times$ bandwidth overhead is a small-cost.

Several high-throughput blockchains have recently emerged, capable of processing 10s or even 100s of thousands of tps~\cite{algorand,conflux,prism-system}, or 10--200 Mbps of transaction data. At such transaction rates, inefficient transaction broadcast mechanisms like flooding become problematic. Simply in terms of cost, wasted data transfers could cost a node-operator tens of thousands of dollars per year.\footnote{For example, a node running in AWS sending 100 Mbps of wasted data at \$0.05/GB~\cite{aws-cost} would incur \$19,710 of extra data transfer cost per year.} Moreover, excessive bandwidth requirements limit the type of nodes that can participate in a blockchain network, hurting decentralization and security. 

A few blockchains have started adopting more efficient broadcasting strategies to address this challenge. A basic approach is to flood {\em hashes} of the transactions, rather than the actual transactions, and have the nodes request only the transactions they are missing from their peers. However, as we discuss in \S\ref{sec:related-works}, it is difficult to ensure both low communication overhead and low latency with such approaches. A node may receive the same hash from multiple peers concurrently and must decide which peers to request the transaction from and how long to wait for a response. For example, to reduce redundant transmissions, Shrec~\cite{shrec} proposes that a node sends only one request for each transaction (hash) and retry the request with a different peer upon a timeout. Unfortunately, there is no good way to set these timeouts in an adversarial environment. A malicious peer can subvert any attempt to tune the timeouts based on delay measurements by arbitrarily injecting jitter to their responses. As a result, practical realization of this idea rely on coarse timeouts (e.g., Conflux uses a 30-second timeout~\cite{conflux-parameters}), which can lead to high broadcast latency even with a small fraction of adversarial nodes. Moreover, hash-based flooding is not effective for small transactions (a common case) where hashes are comparable in size to the actual transactions.

In this paper, we present \name{}, a new approach to transaction broadcasting that achieves both low bandwidth overhead and low latency. We observe that the central issue in existing approaches is that they require {\em coordination} among peers to decide whether to forward individual transaction. For example, hash-based schemes communicate hashes to try to prevent redundant transmissions, but as discussed above performing such coordination efficiently (with low latency) is hard in an uncertain, adversarial setting. An interesting alternative, which we adopt in \name{}, is to use {\em coding}. At a high level, nodes can encode transactions they've received into codewords that they send to their peers, and once a node receives enough codewords ``covering'' a set of transactions, it is able to decode them. Since it does not matter which specific codewords arrive at a node, this approach does not require any coordination, including timeout and retry mechanisms. 

Network coding techniques have been used in variety of contexts for efficient data delivery, including in wireless networks~\cite{katti2006xors}, content distribution networks~\cite{gkantsidis2005network}, peer-to-peer networks~\cite{gkantsidis2006comprehensive, yang2008peer},
multimedia streaming~\cite{magli2013network}, and inter-datacenter transfers~\cite{tseng2021codedbulk} (see~\cite{fragouli2006network} for a review). However. the use of coded transmission raises unique challenges in adversarial blockchain networks. For example, standard techniques such as random linear network coding (RLNC)~\cite{rlnc, algebraic} are susceptible to pollution attacks where a malicious node can send invalid codewords that cause decoding to fail, or worse, propagate through the network corrupting codewords sent by honest nodes~\cite{rlnc-pollution}.

\name{}'s design address two key challenges. First, to safeguard against malicious nodes, we design the encoding and decoding procedures such that a node can efficiently identify and discard invalid codewords during decoding. The key observation is that a certain decoding procedure known as the ``peeling decoder'' enables a node to check the validity of a codeword before using it to decode other codewords. Specifically, \name{} uses LT codes~\cite{lt-codes}, a popular rateless erasure code. LT codes generate codewords by XORing a random subset of transactions. The peeling decoder iteratively removes (``peels off'') previously decoded transactions from received codewords until there's only one transaction left in the codeword. At this time that transaction has been decoded, but before using it to decode other transactions, we can check its validity (see \S\ref{sec:design} for details). Unlike methods that simultaneously decode many codewords together (e.g., solving a system of linear equations to decode RLNCs), an invalid codeword has no effect on decoding other codewords in \name{} (other than a small amount of wasted computation). 

The second challenge is to enable multiple nodes with potentially overlapping transactions to send codewords to a common peer such that they can be {\em jointly} decoded efficiently. A key question here is at what rate should a node send codewords to each peer? The optimal rate depends on the rate at which the other neighbors of that peer send it codewords, and the amount of overlap with the set transactions covered by codewords sent by those peers. For example, if multiple senders have entirely overlapping set of transactions, they can each send fewer codewords per transactions than if they must each communicate a distinct set of transactions. Coordinating the sending rates of multiple nodes by explicitly sharing information about their set of transactions isn't possible\,--\,we'd be back to where we started without coding! 

Our main insight is that explicit coordination isn't necessary for making rate control decisions. Each peer can independently set a codeword rate for each peer to using a simple invariant: {\em send at a rate that achieves a small target loss rate (e.g., 2\%)}, where loss refers codewords that the peer is unable to decode within a certain time period. This invariant ensures that nodes automatically adjust their rates based on how crucial their codewords are to successful decoding at their peers. By targeting a small loss rate, nodes will not send excessively more codewords than needed but send enough to ensure that most codewords can be decoded. As the overlap among the transaction sets of multiple nodes sending to a common peer increases, each will independently reduce its sending rate to maintain the target loss rate, ensuring that redundant transmissions remain low. \name{} uses a simple multiplicative-increase multiplicative-decrease (MIMD) rate control algorithm operating between each pair of peers to continually track the target loss rate.

We evaluate \name{} in both real-world experiment and extensive realistic simulations. Our main findings are:
\begin{itemize}
    \item Coding significantly reduces the communication cost and the latency of transaction broadcasting. \name{} consumes up to 6.38$\times$ less bandwidth while achieving 34\% lower latency than Bitcoin's scheme.
    \item \name{} is resilient to attacks from strong adversaries. An attacker with zero-delay connections to every node in the network can censor a transaction with only 2\% success probability, and increase the latency by 0.1s. In comparison, the latency of Conflux's scheme increases by 15$\times$ when only 4\% of nodes become adversarial.
    \item \name{} scales gracefully to large topologies. As the network grows from 250 nodes to 4000 nodes while maintaining the same degree, the latency of \name increases by 0.5s, while the communication cost increases by 4\%.   
    \item \name{} is computationally efficient. Our implementation can encode or decode 647k transactions per second using one CPU core.  
\end{itemize}

%% file: related-works.tex
\section{Background and Related Work}

A blockchain network consists of many \textit{nodes}. Each node connects to one or more other nodes (\textit{peers}) to form a connected graph. Each node may create new \textit{transactions} (byte strings) that it wishes to write to the blockchain.%
The nodes use a {\em transaction broadcast} mechanism to spread their transactions throughout the network. The goal of transaction broadcasting is to deliver new transactions to most nodes in the network with low latency and low communication overhead. Here, latency refers to the time span from when the first node sends this transaction to the time that a large fraction (e.g., 95\%) of nodes have received that transaction. The communication overhead is the average amount of data that each node downloads relative to the transactions it receives. (We will define these performance metrics more precisely for evaluation, in \S\ref{sec:eval}.) A crucial aspect of a blockchain's peer to peer network is resisting adversarial nodes. Such nodes may deviate from the specified communication protocol arbitrarily to disrupt transaction broadcasting (e.g., to censor or delay transactions).

\subsection{Existing mechanisms}\label{sec:related-works}

\smallskip\noindent\textbf{Transaction flooding.} The simplest broadcast mechanism is transaction flooding, where nodes forward new transactions to all peers (except the sender) as soon as they receive a new transaction. Transaction flooding incurs a high overhead, because a node will receive every new transaction as many times as the number of peers it has, since each peer will forward the transaction upon receiving it. On the other hand, new transactions travel along the shortest path from their creators to every node, so the latency is optimal. The scheme is also robust against attacks. As long as there is a path between every pair of honest (i.e., non-adversarial) nodes that does not contain any adversarial node, new transactions can be delivered to all honest nodes. A generalization of flooding is to have each node forward transactions only to a random subset of peers. This reduces the overhead but increases the latency (transactions may not travel along the shortest paths) and makes the system more vulnerable to attacks (nodes may accidentally forward a transaction to only adversarial nodes). Ethereum uses this method~\cite{ethereum-measure}. 

\smallskip\noindent\textbf{Hash flooding.} Instead of flooding the actual transactions, an alternative is to only flood their \textit{hashes}. Upon receiving a hash from a peer, a node checks if it has already obtained the corresponding transaction (i.e., the preimage). If not, it requests for the preimage from the peer. After receiving the preimage, the node forwards the hash to all peers. This scheme reduces the communication overhead at the cost of an extra round-trip of latency at each forwarding hop. 

Hash flooding, however, does not eliminate all redundant communication. A node may hear about the same transaction hash from multiple peers concurrently, and then has to decide which peers to request the transaction from. One option is to request it from all peers that forward the hash but this approach incurs a high communication overhead. Shrec~\cite{shrec} proposes to mitigate this issue by ensuring that a node only sends one request for each transaction (hash), so that at most one peer will reply with the preimage. However, this modification is not robust to attacks. If a request goes to an adversarial peer, it may not respond with the preimage, or may deliberately delay the response. Shrec does not specify a solution, and nodes in actual deployments rely on heuristics (such as waiting for a long time that bounds the network latency) to decide whether to keep waiting or retry requests with another peer~\cite{conflux-parameters}. These heuristics can lead to high latency in the presence of a small fraction of adversarial nodes (as we show later in experiments).
Another mitigation proposed by Bitcoin is to require each node to inject a random jitter before forwarding a hash to peers, so that nodes are less likely to receive the same hash from multiple peers within a short time window. However, this mitigation is only effective when the jitter is sufficiently long, and forces a trade off between latency and communication overhead (as we show later through experiments as well).

Although hashes are shorter than the actual transactions, the cost to flood them is still significant due to the high overhead of flooding. For example, hash flooding accounts for 30--50\% of all the network traffic of a Bitcoin node~\cite{erlay}.
Simply reducing the length of the hashes allows adversarial nodes to produce hash collisions and disrupt the system. (Recall that nodes will not request a transaction if they have already received another one with a colliding hash.) Bitcoin~\cite{compact-block} and Shrec~\cite{shrec} propose an elegant solution by using a keyed hash, such as SipHash~\cite{siphash}. Each pair of connected nodes use a local secret key when computing hashes. Adversarial nodes cannot produce hash collisions without knowing the secret key.
Other works such as Graphene~\cite{graphene} and Erlay~\cite{erlay} propagate hashes using \textit{set reconciliation}~\cite{set-reconciliation}, a primitive that allows two nodes to exchange hashes with optimal overhead. However, it only changes how nodes propagate hashes. They still need to decide whether to request for a transaction from a peer after it receives a hash through set reconciliation, so they suffer from the same issue as in the hash flooding schemes. Moreover, experiments in a recent work~\cite{shrec} show that set reconciliation is impractical for high-throughput blockchains due to high computational cost and high sensitivity to parameter choices, facing a computational bottleneck at 826 tps on a 4-core server. %

\smallskip\noindent\textbf{Structured propagation.} The peer-to-peer networking literature has proposed a variety of efficient broadcast mechanisms~\cite{chord-broadcast, splitstream, coopnet} that forward requests on a structured topology (e.g., a broadcast tree). These methods were not adopted by blockchain systems largely because they are  not tolerant to Byzantine nodes. For example, adversarial nodes at higher levels of a broadcast tree can disconnect it and thwart the progress of the broadcast~\cite{poison-kad}.

%% file: overview.tex
\section{The Case for Coding and its Challenges}

\label{sec:motivation}

Nodes in existing transaction broadcasting mechanisms store and forward \textit{individual} transactions. This design creates a common challenge: how to avoid forwarding redundant transactions to peers? To see why it is difficult, let us consider a node $A$ with two peers $P_1$, $P_2$. Suppose that at some point, $P_1$ receives a new transaction $t$ from the rest of the network and needs to decide whether to forward it to $A$. Specifically, $P_1$ needs to know whether $A$ has already obtained $t$ from $P_2$ in order to avoid forwarding redundant data. The previous section described a few strategies current systems have proposed to obtain this information (e.g., hash flooding and explicit transaction requests). However, as we described, \textit{uncertainty} about network delays and the adversarial nodes makes it difficult, if not impossible, for $P_1$ to consistently make an optimal decision on whether to forward $t$ or not using such strategies. When $P_1$ makes a mistake, it either increases the overhead (forwarding a redundant transaction), or increases the latency (not forwarding a transaction that $A$ does not have). To make the matter worse, nodes in blockchain networks connect to many peers in order to minimize the risk of being disconnected~\cite{bitcoin-eclipse}. In reality, $A$ would have much more than two peers, e.g., 8--133 in Bitcoin~\cite{erlay} or 25--50 in Ethereum~\cite{ethereum-measure}. Each of the peers is prone to making such a mistake and increasing the overhead or the latency, and the impact from each peer accumulates. This happens for every new transaction in the system.

\subsection{\name: coded transaction broadcast}
Communication theory suggests that we can solve the above problem with network coding~\cite{nc}. Instead of forwarding individual transactions, nodes compute and forward linear combinations of multiple transactions (treating them as elements of a finite field), along with the coefficients for computing them. We call each linear combination and the corresponding coefficients a \textit{codeword}. The node receiving a codeword treats it as a linear equation, where the transactions are the unknowns. By collecting enough linearly independent equations (codewords), the node can solve the system of equations and recover the transactions.

The key benefit of coding is that it eliminates the need for nodes to make optimal decisions on whether to forward each individual transaction, which we have shown to be error-prone. For example, suppose that $P_1$ receives two new transactions $t_1$, $t_2$, but $A$ has already obtained $t_1$. Plain store-and-forward would have required $P_1$ to decide whether to forward $t_1$ and $t_2$ respectively, each prone to incurring overhead or latency. Instead, $P_1$ in network coding may compute $t_1 + t_2$ and send the resulting codeword to $A$.\footnote{In reality, $P_1$ should randomize the coefficients so that the resulting codeword is highly likely to be linearly independent to other linear combinations of $t_1$ and $t_2$~\cite{rlnc}. We omit this step to simplify the example.} $A$ can recover $t_1$ by computing $(t_1+t_2)-t_2$. Notice that $P_1$ does not need to decide \textit{which} transaction to forward, because the codeword $t_1+t_2$ is useful to $A$ no matter which one of $t_1$ and $t_2$ it is missing. Instead, $P_1$ only needs to decide \textit{how many} codewords to send (in this simple case, one). This task is much easier, because $P_1$ does not need precise information about which transactions $A$ has already received from other peers.

\smallskip \noindent
{\bf Coding in an adversarial network.}
While network coding holds great promise, it is difficult to implement in networks with adversarial nodes. Recall that a codeword is a linear combination of multiple transactions along with the corresopnding coefficients. The main challenge is to defend against \textit{invalid} codewords, where the linear combinations are inconsistent with the stated coefficients. When a node puts an invalid codeword into a linear system and tries to solve it, it either obtains corrupt solutions (transactions), or the linear system is inconsistent and unsolvable. The node cannot tell which equations are invalid before solving the linear system, because it must obtain all the involved transactions and recompute the linear combinations locally. As a result, the node has to enumerate all subsets of possible equations (codewords) to include in the linear system until it finds one that is consistent, which is costly since there are exponentially many subsets of codewords to consider
Furthermore, standard network coding allows nodes to produce new codewords by recombining existing ones~\cite{nc}. This can amplify the impact of invalid codewords since honest nodes may unknowingly use them when producing new codewords, which will in turn become invalid and impact other honest nodes in the same way.

To enable coding in an adversarial setting, we need a way to identify invalid codewords and limit their impact on the decoding process. Our main insight is that codes that can be decoded via a ``peeling decoder'' are suitable for this purpose. The peeling decoder decodes a transaction from a codeword only using transactions it has already decoded. This allows the decoder to check the validity of each transaction it decodes before using it in the decoding process, thus preventing any corrupt codeword from interfering with other transactions.

Specifically, \name{} builds on LT codes~\cite{lt-codes}, a popular rateless erasure code in our construction. 
LT codes generate codewords by XORing a number of randomly-chosen transactions. The number of transactions varies from codeword to codeword, and is itself chosen at random according to a carefully-chosen {\em degree distribution}. During decoding, LT codes iteratively ``peel off'' previously decoded transactions from codewords in which they appear. For example, suppose codeword $C = T_1 \oplus T_3 \oplus T_7$ is the XOR-sum of three transactions. When the decoder decodes transaction $T_1$, it removes it from $C$ by computing $C\oplus T_1 = T_3 \oplus T_7$. Whenever a codeword is left with only one transaction after a peeling step, that transaction has been decoded. However, before using it to decode any other codewords, the decoder can check its validity. \S\ref{sec:design:ids} describes how \name{} performs this validity check by including a short hash of the transactions in each codeword, which also serve as transaction IDs needed by the peeling decoder.

\noindent {\bf Rate control.} A second key challenge for coded broadcast is for nodes to decide the rate to send codewords to each peer. The crux of the issue is that the neighbors of a specific node may have partially overlapping sets of transactions that must be delivered to their common peer. To minimize overhead, the nodes should send fewer codewords per transaction when their transaction sets overlap significantly and more codewords when their sets are distinct (note that the receiver decodes codewords received from all peers jointly). But the nodes don't know how much overlap exists between their transactions or how other nodes will behave. \name{} uses a dynamic rate control algorithm running  between every pair of peers (\S\ref{sec:design-controller}) to automatically adapt the sending rates and discover an efficient operating point. \name{}'s rate control is robust to adversarial nodes, since each node independently makes its own rate control decisions. A malicious node deviating from the rate control protocol can waste bandwidth locally but cannot prevent the progress of decoding at honest nodes.\footnote{An adversarial node sending excessive amounts of traffic can be identified and disconnected using standard techniques.}

%% file: design2.tex
\section{Design}
\label{sec:design}

\name{} runs on a peer-to-peer network, where each node connects to several peers to form a connected graph. Nodes can create new transactions, produce codewords from transactions they create or receive from other nodes, and learn of new transactions by decoding codewords they receive from their peers. 
In this section, we explain the design of \name in four steps.  \S\ref{sec:design-encoder-decoder} describes the basics of how nodes encode and decode codewords. Here, we make a simplifying assumption that all transactions have equal length, which we will remove later. \S\ref{sec:design-multi-peers} describes how nodes control the rate for sending codewords to their peers. \S\ref{sec:design-byzantine} discusses how \name defends against adversarial nodes by making codewords self-validating. \S\ref{sec:design-multi-fragment} removes the prior assumption on fixed transaction length and adds support for variable-sized transactions.

\subsection{Windowed LT codes}
\label{sec:design-encoder-decoder}

\name uses a modified version of LT codes~\cite{lt-codes} to encode transactions, 
where each codeword consists of the bit-wise XOR sum of one or multiple transactions and the list of their identifiers. Standard LT codes are designed to encode a finite fixed block of data symbols. In comparison, nodes in \name receive and create transactions continuously, and hence have to encode an infinite stream of transactions. While it is possible for nodes to buffer and partition the streams into finite, disjoint segments of transactions, and encode each segment independently using conventional LT codes, this process incurs undesired latency since the encoder will need to accumulate enough transactions to fill its buffer before issuing codewords. Instead, we modify LT codes to encode streams of transactions directly. Our approach, which we refer to as \textit{windowed} LT codes, is to pass the stream of transactions through a first-in-first-out (FIFO) coding window, and encode the transactions in the window using an LT code.

\smallskip\noindent\textbf{Encoding.} Each node maintains a FIFO \textit{coding window} which contains the $k$ most-recent transactions it receives or creates. The process for generating a codeword is the same as in conventional LT codes, except that the node only considers the transactions inside the window. Specifically, the node randomly samples a number $d \in \{1, \ldots, k\}$ from a discrete probability distribution $\rho$. We call $d$ the \textit{degree} of the codeword, and $\rho$ the \textit{degree distribution} of the code. It then uniformly samples $d$ transactions from the coding window as the \textit{source transactions} for the codeword. The resulting codeword contains a list of identifiers for the source transactions in the codeword and their bit-wise XOR sum. We call them the \textit{header} and the \textit{payload} of the codeword, respectively. The identifiers in the header let the decoding node determine the set of source transactions (\S\ref{sec:design:ids} describes the identfiers in detail). For now, we assume that all transactions have the same length, so that the bit-wise XOR operation is well defined. %

\smallskip\noindent\textbf{Decoding.} To decode codewords, nodes run an iterative \textit{peeling} algorithm as in conventional LT codes. Each node maintains a bipartite graph of undecoded codewords and transactions. Specifically, there is an edge between a codeword and a transaction in the bipartite graph if and only if the latter is a source transaction of the former. (The node determines such relationships by reading the identifiers in codeword headers.) When a codeword has only one neighbor, the payload of that codeword must be the XOR sum of one transaction, i.e., the neighbor transaction itself. The node can therefore decode such ``degree-one'' codewords immediately. Having decoded a transaction, the node  ``peels'' it from its other neighboring codewords by XORing it into their payloads, and then deletes the transaction and the codeword from the graph. Upon receiving a new codeword, the node checks for any source transactions it has already decoded, and peels them before inserting the codeword and its remaining (undecoded) source transaction into the graph.

We would like to minimize the \textit{overhead}, i.e., the average number of codewords that the decoder must receive to decode a transaction. Therefore, ideally, removing a newly-decoded transaction from the bipartite graph should cause more codewords to become decodable (i.e., have exactly one neighbor), so that the decoding process can continue without receiving more codewords. Prior work shows that one can design the codeword degree distribution $\rho$ such that the peeling algorithm can recover a block of $k$ data symbols from any $k + O(\sqrt{k} \ln^2(k/\delta))$ codewords with high probability (at least $1-\delta$)~\cite{lt-codes}. The Robust Soliton distribution~\cite{lt-codes} is one such degree distribution that prior work~\cite{lt-codes, raptor-codes} shows can (asymptotically) achieve the optimal overhead in the setting where all codewords are created from the same fixed set of data symbols. 

A windowed LT code applies an LT code on a moving window of transactions, so different codewords may correspond to different sets of transactions as the content of the window changes. However, as shown in \autoref{fig:windowed-lt-comparison}, windowing has negligible effect on the overhead of the code. The figure shows the overhead for a conventional LT code transmitting a fixed block of $k$ transactions and a windowed LT code with window size $k$ when using the same degree distribution. Intuitively, at any given time, the peeling process for a windowed LT code with window size $k$ is similar to the peeling process for a fixed set of $k$ transactions. Hence, the number of codewords required to successfully decode a window's worth of transactions in a windowed LT code is, on average, the same as the number of codewords necessary in a conventional LT code for $k$ (fixed) transactions. Though we leave a formal analysis to future work, the decoding process of windowed LT codes can be analyzed rigorously using the spatial-coupling technique~\cite{spatial-coupling} in coding theory. Specifically, windowed LT codes are the limit of spatially-coupled LT codes~\cite{coupled-lt} as the length of the encoded data goes to infinity. Theoretical and empirical results~\cite{spatial-coupling, spatial-coupling-2, coupled-lt} show that spatially-coupled codes achieve equal or lower overhead than their uncoupled counterparts, and the results hold for LT codes~\cite{coupled-lt}.

\subsection{Codeword rate controller}
\label{sec:design-multi-peers}
\label{sec:design-controller}

\begin{figure*}
\centering
\begin{minipage}[t]{.33\textwidth}
  \centering
  \captionsetup{width=.95\columnwidth}
  \includegraphics[width=\columnwidth]{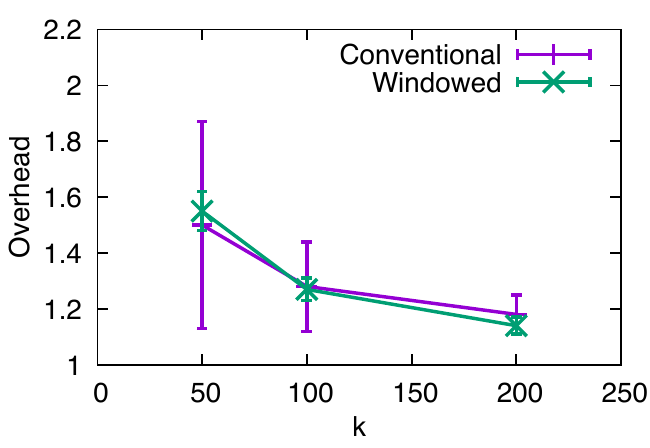}
\caption{Overhead of conventional and windowed LT codes for different $k$. The LT encoder uses the Robust Soliton distribution with $c=0.03$, $\delta=0.5$. See \cite{lt-codes} for definitions of $c$ and $\delta$.}
\label{fig:windowed-lt-comparison}
\end{minipage}%
\begin{minipage}[t]{.33\textwidth}
  \centering
  \captionsetup{width=.95\columnwidth}
  \includegraphics[width=\columnwidth]{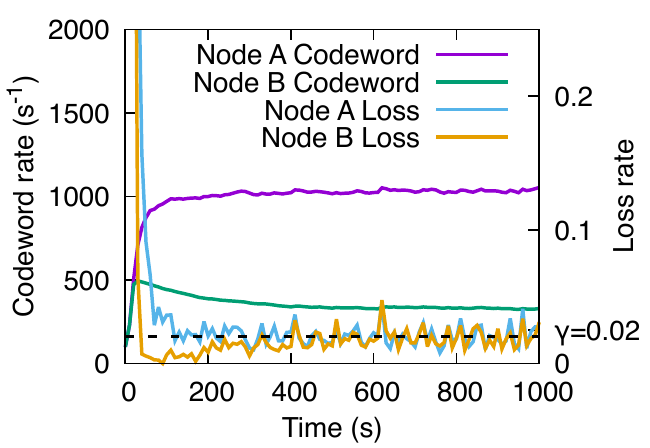}
\caption{Time series of the codewords rates and the loss rates of two nodes $A$, $B$ sending to a common peer. The dashed line shows the target loss rate $\gamma=0.02$. The set up is described in \S\ref{sec:design-controller}.}
\label{fig:controller-timeseries}
\end{minipage}%
\begin{minipage}[t]{.33\textwidth}
  \centering
  \captionsetup{width=.95\columnwidth}
  \includegraphics[width=\columnwidth]{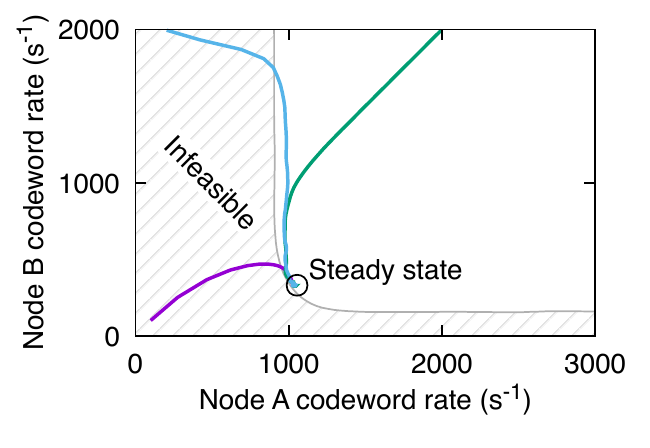}
\caption{Transition of the codeword rates of two nodes $A$, $B$ sending to a common peer. Each curve corresponds to a different initialization of codeword rates. The set up is described in \S\ref{sec:design-controller}.}
\label{fig:controller-trace}
\end{minipage}
\end{figure*}

As described in \S\ref{sec:design-encoder-decoder}, nodes do not need to decide \textit{which} transactions to include in a codeword. However, they still need to determine \textit{how fast} they should send codewords to each peer. This is critical to the reliability and the efficiency of the system. The rate should be high enough to ensure quick decoding of most transactions, while as low as possible to minimize the communication cost. %

In \name, nodes target a constant \textit{loss rate} for the codewords they send to each peer. Specifically, a node tries to maintain that each peer decodes $(1-\gamma)$-fraction of its codewords within time $\tau$ of receiving each codeword. We say a \textit{loss event} occurs when the peer fails to decode a codeword in time. We call $\tau$ the \textit{decoding timeout}, and $\gamma$ the \textit{target loss rate} ($0 < \gamma < 1$). 
Nodes do not target a zero loss rate; trying to do so would require sending many redundant codewords to guarantee absolutely all transactions can be decoded~\cite{raptor-codes}.
Instead, nodes target a small but non-zero loss rate, and there are two reasons. First, sparse loss events are useful signals that indicate a node is sending just enough codewords for a peer to successfully decode. Second, a low loss rate has only minimal impact on a node's ability to receive transactions (since the transactions covered in those codewords are also likely to be covered in other codewords). In our implementation (\S\ref{sec:eval}), we target a loss rate of 2\%, and every node consistently receives more than 95\% of transactions. %

To maintain the target loss rate $\gamma$, nodes run a multiplicative-increase multiplicative-decrease (MIMD) algorithm to control the rate they send codewords to each peer. Besides $\gamma$ and $\tau$, the algorithm has a parameter $\alpha$ that controls its aggressiveness. Let $r$ be the current rate that a node sends codewords to a peer. Every time the node sends a new codeword to the peer, it decreases the rate by $r\alpha\gamma$. Every time the peer reports a loss event (with respect to timeout $\tau$), it increases the rate by $r \alpha$. To see the behavior of this algorithm, consider a short time span $\Delta$. The node sends $r\Delta$ codewords to the peer, causing the controller to decrease the rate by $r\Delta\alpha\gamma$. Meanwhile, let $P_\text{loss}$ be the current loss rate. During the same period, there are $r\Delta P_\text{loss}$ loss events, causing the controller to increase the codeword rate by $r\Delta P_\text{loss} \alpha$. After the time span $\Delta$, the codeword rate becomes $r[1+\Delta \alpha(P_\text{loss}-\gamma)]$, a multiplicative increase/decrease of factor $1+\Delta \alpha(P_\text{loss}-\gamma)$. When $P_\text{loss} > \gamma$, the algorithm causes the node to increase the codeword sending rate, making it easier for the peer to decode (because it will receive more codewords) and thus reducing $P_\text{loss}$. The opposite happens when $P_\text{loss} < \gamma$. Either way, the algorithm drives $P_\text{loss}$ to $\gamma$, which is a steady state where the codeword rate changes by a factor of 1, i.e., no change. When $\alpha$ is sufficiently small (e.g., 0.1 in our implementation), the algorithm converges to $P_{loss} = \gamma$.

The MIMD algorithm enforces the target loss rate $\gamma$ for every pair of peers when decoding each other's codewords. This key invariant ensures that nodes send codewords at an appropriate rate to each peer regardless of the topology or the transaction pattern at different nodes. Intuitively, nodes that receive more \textit{novel} transactions (which peers do not have) will send codewords at a higher rate to their peers in order to maintain the target loss rate, and the opposite goes for nodes with fewer novel transactions. To demonstrate this behavior, let us walk through a simple example, where two nodes $A$, $B$ connect to a common peer. $A$ receives unique transactions (from the outside) at 600 tps, and $B$ receives unique transactions at 100 tps. In addition, both nodes receive a common flow of transactions at 400 tps. $A$ and $B$ both send codewords to the common peer. \autoref{fig:controller-timeseries} shows the time series of the codeword rates from $A$, $B$ as the control algorithm adjusts them, and the loss rates of their respective codewords at the peer. Notice that after the controller converges, $A$ sends codewords at a higher rate than $B$, because it has more novel transactions. The loss rates of the codewords from both nodes stabilize at 0.02, which is our control target $\gamma$. \autoref{fig:controller-trace} shows the transition of the codewords rates as we rerun the example with three different initial rate configurations for $A$ and $B$. Regardless of the initial state, their codeword rates converge to the same point. In the same figure, we plot the infeasible region, which are codeword rates that will cause a higher loss rate than $\gamma$ (for codewords from either $A$ or $B$). Notice that the controller converges right on boundary of the infeasible region, indicating that it accurately achieves the target loss rate $\gamma$. Moreover, it converges at the exact point on the boundary where the \textit{sum} of the codeword rates from $A$ and $B$ is minimized, which means the total communication overhead is optimal.

Note that nothing prevents an adversarial node from deviating the control algorithm. Specifically, it may ignore the feedback and send at an excessively low or high rate. A low rate would not have much affect, since the other (honest) peers would compensate by increasing their rates. A high rate can waste the peer's bandwidth. Similar to any denial of service attack, we could use standard mechanisms to identify and disconnect from such nodes. In fact, the receiver can predict the expected sending rate based on its feedback, so it is easy to detect outliers. Finally, a malicious node could lie about loss events to its peers, causing them to increase their rates. However, the effect is bounded, since a node can easily determine an upper limit on the codeword rate necessary to decode its transactions, assuming the peer only receives from this node.

\subsection{Transaction identifiers}
\label{sec:design-byzantine}
\label{sec:design:ids}
As mentioned in \S\ref{sec:design-encoder-decoder}, codewords include the identifiers of the source transactions, so that the peeling algorithm can decide whether to peel a transaction off a codeword. Nodes in \name{} use codewords they receive from all their peers to decode transactions. This reduces the communication overhead since the decoder can consider information received from all peers collectively. However, it requires a consistent way to identify transactions across peers, so that a transaction decoded from one peer's codeword can be used to decode a codeword from another peer. Specifically, the decoder needs to compute the identifier of a transaction based on the transaction's data (which is consistent across all peers). A crucial challenge induced by this approach is that blockchain transactions are often rather short (e.g., the smallest Ethereum transaction is only 109 bytes, which is also the most prominent size~\cite{eth-bigquery}). Therefore, using long hashes that ensure collision-free IDs (e.g., 32B long for SHA256) incurs relatively high overhead. Instead, \name uses a short hash to identify transactions in codewords and its decoder handles collisions that may occur.

\smallskip\noindent\textbf{Detecting hash collisions.} 
Hash collision affects the decoding process by causing the decoder to incorrectly peel a transaction off a codeword. Consider three different transactions $T_1, T_2, T_3$. There is a hash collision between $T_1$ and $T_2$, i.e., $H(T_1) = H(T_2)$. Suppose the decoder has previously decoded $T_1$, and now receives a new codeword whose source transactions are $T_2$ and $T_3$. The decoder sees that $H(T_1)$ appears in the header (which is actually $H(T_2)$, but they are equal), and mistakenly peels $T_1$ off the codeword. The header becomes $\{H(T_3)\}$, so the decoder believes that the codeword is now degree-1, and treats the payload as a newly-decoded transaction. However, the payload is actually $T_1+T_2+T_3$, and is not a valid transaction.

The example above already hints at a solution: it is highly likely that $H(T_1+T_2+T_3) \ne H(T_3)$. Before taking the payload from a degree-1 codeword, the decoder should compute its hash and compare it with the hash in the header. If the hashes do not match, the decoder knows that the codeword is corrupted, and throws it away. Although a corrupted codeword might still pass the check by chance, e.g., when $H(T_1+T_2+T_3) = H(T_3)$, it is simply another case of hash collision, and can be handled in the same way. Also, the possibility that the resulting invalid transaction causes further hash collisions decays exponentially (and can be handled in the same way as we just described if it does happen).  

\smallskip\noindent\textbf{Minimizing hash collisions.} We now have a mechanism to detect hash collisions. The
decoder discards the affected codewords and recovers from collisions at the cost of some bandwidth waste. However, hash collisions become more frequent as the system runs, because there are more decoded transactions to collide with the source transactions in a new codeword. As an extreme example, after sufficiently long time, the decoder will have decoded a transaction for every possible hash value. At that point, every new codeword will experience hash collision, and the decoder may stall.

\name addresses this problem by having nodes to only use the last $m$ transactions they receive to decode the current codeword. We call $m$ the size of the \textit{peeling window}. $m$ controls the probability of hash collisions. Specifically, let $h$ be the length of the hashes in bits. Assuming that hashes are uniformly distributed, the probability that a new transaction collides with an existing transaction in the peeling window is $m/2^h$. Because the maximum degree of a codeword is $k$, the probability that a codeword is corrupted due to hash collision is at most $k\cdot m/2^h$ by union bound. For example, using $m=100000$ and $h=4 $ bytes, the upper bound on the corruption probability is 0.0023\%. Recall that the codeword rate controller already targets a non-zero loss rate, e.g., 2\% in our implementation. The extra loss from corrupted codewords is negligible.

\smallskip\noindent\textbf{Adversarial hash collisions.}
We have discussed how the decoder mitigates hash collisions when they appear at random, i.e., not deliberately introduced by an adversary. However, when the hashes are short, an adversarial peer can generate collisions. Given a transaction ID (hash image), the attacker can enumerate transactions until finding one that hashes to the same ID. Such an adversary could disrupt communication in the peer to peer network by broadcasting transactions with colliding IDs to legitimate transactions. For example, the attacker may attempt to create collisions with Alice's transaction in order to censor it. 

The solution is to prevent the adversary from computing the hash function that a node uses, so that it cannot forge hash collisions to affect the node. \name achieves this by leveraging a \textit{keyed} hash function to set the transaction ID, which depends on both the transaction data and a \textit{secret key}. This design is inspired by Short Transaction ID in Bitcoin~\cite{compact-block}. A node sends a fresh secret key to each of its peers. That peer will then use this secret key to compute the transaction IDs in the codewords it sends to the node. Whenever the node wants to use a transaction it recovers from one peer to decode a codeword from another peer, it re-hashes this transaction using the second peer's key to obtain its ID relative to that peer. This technique is also enabled by the peeling algorithm, which relies on transactions that are already decoded to decode new transactions from codewords.

Using a keyed hash, the attacker cannot create a transaction with a hash ID that collides with another specific transaction, since such collisions are peer-dependent. This prevents the adversary from, e.g., censoring a transaction from Alice. It ensures that a node that connects to good peers will be able to decode that transaction despite being also connected to rogue peers.

\subsection{Variable-sized transactions}
\label{sec:design-multi-fragment}
\name nodes generate codewords by computing the bit-wise XOR sum of the source transactions.
So far, we have assumed that transactions are of the same size, so that the sum has the same length as each source transaction (operand). However, when some transactions are longer than others, the resulting bit-wise XOR sum is as long as the largest operand. This causes inefficiency, because the size of a codeword's payload needs to accommodate the largest source transaction, while the codeword may encode many smaller transactions. 

Instead, \name fragments large transactions. The key challenge is that an adversary may attack the reassembly process by sending conflicting fragments. To see the potential problem, consider a straw man solution where each fragment contains a sequence number and an identifier (e.g., hash) of the transaction it belongs to. It is impossible for a node to validate these two fields, unless it has received the whole transaction, in which case it does not need the fragment anymore. As a result, an adversary can create fake fragments for a transaction. Although transactions usually contain signatures so their integrity is never at risk, fake fragments greatly increase the computational cost for reassembling a transaction, because nodes must enumerate an exponential number of combinations of (authentic and fake) fragments to correctly reassemble it.

In \name, nodes fragment long transactions to fixed-size fragments of $\ell$ bytes. To defend against conflicting fragments from the adversary, each fragment contains the cryptographic hash of the previous fragment of the same transaction. The hash ensures that there is only one unique prefix (ending with the fragment identified by the hash) that this fragment should attach to, so that there is no ambiguity during reassembly. Each fragment also contains two marker bits that indicate whether it is the first or the last fragment in the transaction. The remainder of the $\ell$ bytes include data from the transaction. Fragments are broadcast in the system in the same way as fixed-size transactions. When a node receives the last fragment of a transaction (it can tell by examining the marker bits), it starts reassembling by tracing the the hash chain until it finds the first fragment (also by examining the marker bits). It then concatenates the data in each fragment to form the transaction.

Here, the fragment size $\ell$ decides the maximum transaction size that the system can support without fragmentation, and the relative overhead of the hashes (because every $\ell$-byte fragment contains a hash). Finding the optimal $\ell$ given a distribution of the transaction size is a simple optimization process, because the total size of all fragments of a transaction is determined by $\ell$ and the size of the transaction. We expect fragmentation to only introduce minimal overhead, because the size of blockchain transactions is highly concentrated around one or a few values (usually corresponding to the size of simple payments)~\cite{eth-bigquery, btc-bigquery}. For example, we analyzed historic Bitcoin transaction data~\cite{btc-bigquery} from Jan 2019 to Feb 2022. In the optimal setting ($\ell=258$ bytes), the fragmentation overhead is only $1.23\times$, much lower compared to the overhead of the base system ($1.7\times$ to $2.2\times$).

\begin{table}
\centering
\small
\begin{tabular}{ r|l|p{24mm}|c } 
 \hline
 Sym. & Name & In evaluation & Def. \\\hline\hline
 $\rho$ & Degree distribution & Robust Soliton,\newline $\delta=0.5$, $c=0.03$ & \multirow{2}{*}{\S\ref{sec:design-encoder-decoder}} \\
 $k$ & Coding window size & 50 &   \\\hline
 $\tau$ & Decoding timeout & 0.5 ms & \multirow{3}{*}{\S\ref{sec:design-controller}} \\
 $\gamma$ & Target loss rate & 0.02 & \\
 $\alpha$ & Aggressiveness & 0.1 & \\\hline
 $\ell$ & Fragment size & 128 bytes & \multirow{1}{*}{\S\ref{sec:design-multi-fragment}} \\
 \hline
\end{tabular}
\caption{Main parameters of \name and the values used in the evaluation.}\label{tab:parameters}
\end{table}

%% file: evaluation.tex
\section{Evaluation}
\label{sec:eval}
Our evaluation will demonstrate:
\begin{itemize}
    \item \name achieves a better trade off among latency, overhead, and robustness than the state of the art.~(\S\ref{sec:eval-comparison})
    \item \name is resilient to censorship from a strong adversary.~(\S\ref{sec:eval-censorship})
    \item \name can scale to a large number of nodes while maintaining its performance.~(\S\ref{sec:eval-scalability})
    \item The control algorithm in \name adapts to changing transaction load, and converges quickly.~(\S\ref{sec:eval-converge})
    \item Encoding and decoding are computationally efficient, and scales to a throughput on the order of $10^{5}$ tps on commodity computers.~(\S\ref{sec:eval-benchmark})
\end{itemize}

\subsection{Experimental setup}\label{sec:exp-setup}

\noindent\textbf{Implementation and internet testbed.} We implement \name in 1300 lines of Go. To evaluate its performance when running across the internet, we set up a real-world testbed on Vultr\footnote{\url{https://www.vultr.com}} consisting of 19 virtual machines in 19 different cities around the world. Each virtual machine has 4 CPU cores and 8GB of RAM, and hosts one node. Each node connects to 4 random peers, forming a random regular graph of degree 4 and diameter 4.

\smallskip\noindent\textbf{Simulator and topologies.} Two factors limit the size of our internet testbed: (i) the number of datacenter locations that cloud vendors offer; (ii) the high cost to operate a large test deployment across the internet. To experiment with larger topologies, we implement \name as a module for the OMNET++~\cite{omnet} discrete-event simulator in 900 lines of C++. The simulator models the network propagation latency between each pair of nodes. We experiment with three classes of topologies: (i) a replica of the aforementioned testbed, parameterized with latency measurements from the real world; (ii) a degree-16 random regular graph of 250 nodes at 250 cities around the world, parameterized with a public data set~\cite{pingstats} of latency measurements across these cities; (iii) graphs of up to 4000 nodes with pair-wise latency sampled from a log-normal distribution (more details in \S\ref{sec:eval-scalability}). In \S\ref{sec:validate-simulator}, we compare the results from the real-world testbed with the simulator using the same topology, and demonstrate that the simulator is highly accurate.  

\smallskip\noindent\textbf{Parameters.} See \autoref{tab:parameters}.

\smallskip\noindent\textbf{Workload.} To generate workload, nodes create new transactions in Poisson processes with adjustable rates. \name serves as the networking layer for blockchains, orthogonal to the blockchain applications running on top. So, we let each transaction be a random 128B string to match standard blockchain applications~\cite{algorand-tx-type, eth-bigquery}. Transactions do not have blockchain-specific semantics such as payments, smart contracts, signatures, etc.

\smallskip\noindent\textbf{Metrics.} For each node in the system, we measure the following three metrics:
\begin{itemize}
    \item \textit{Latency.} The average latency of all transactions that the node receives. The latency of a transaction is the time from its creation to its reception at the node.
    \item \textit{Delivery rate.} The number of transactions that the node receives, divided by the number of all transactions created in the system. It measures how reliably the node can receive transactions.
    \item \textit{Overhead.} The amount of codeword data that the node downloads, divided by the amount of transaction data it decodes. It is the relative cost compared to a perfect delivery protocol where each bit a node receives translates to a new bit of useful transaction data (1 is perfect in this metric). 
\end{itemize}
A public blockchain usually involves many operators (individuals or companies), each operating one or a few nodes. In order to benefit all the operators, the transaction broadcasting scheme should provide good performance even for nodes at the tail. As a result, unless otherwise mentioned, we report the $95^\text{th}$-percentile across all nodes for each metric. For each datapoint, we collect data over a duration of at least 100 seconds in simulation or real time. 

\subsection{Real-world experiments}
\label{sec:validate-simulator}

\begin{figure}
\centering
\includegraphics[width=\columnwidth]{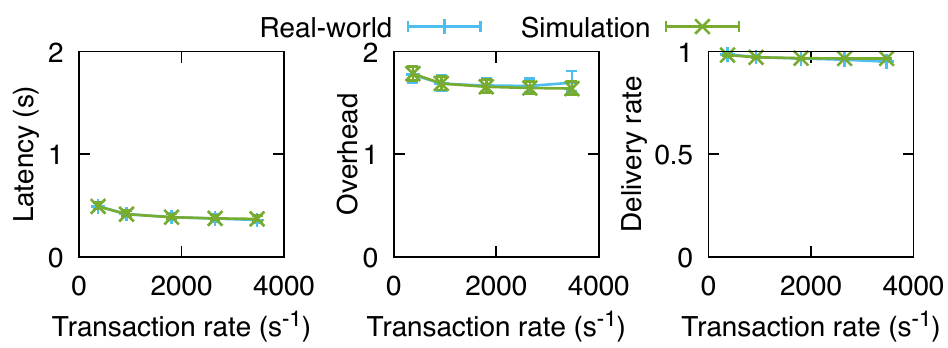}
\caption{Latency, overhead, and delivery rate when running our implementation on the real-world testbed, compared to simulation of the same topology. Notice that the curves overlap. Points show the average, and error bars show the $5^\text{th}$ and the $95^\text{th}$-percentiles across all nodes.}
\label{fig:sim-real-metrics}
\end{figure}

\begin{figure}
\centering
\includegraphics[width=\columnwidth]{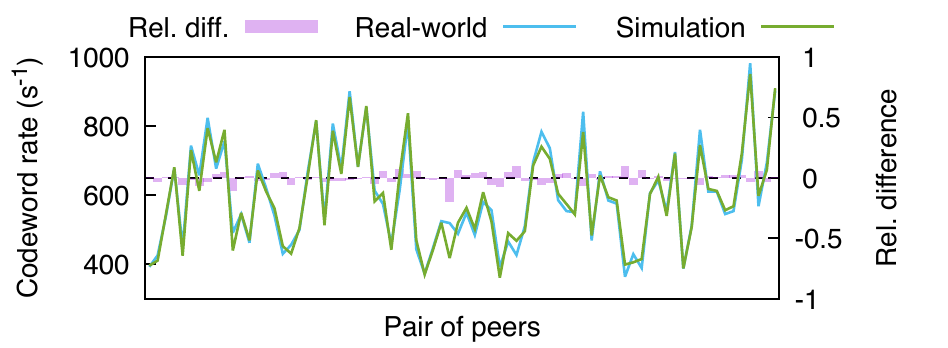}
\caption{The rate that nodes send codewords between each pair of peers when running our implementation on the real-world testbed, compared to simulation of the same topology. Notice that the curves overlap. Bars in the background show the relative difference of the simulation compared to the real world.}
\label{fig:sim-real-edges}
\end{figure}

We first evaluate \name on the real-world testbed. \autoref{fig:sim-real-metrics} shows the performance metrics as we vary the arrival rate of new transactions from 370~tps to 3500~tps. In all scenarios, \name consistently achieves a delivery rate of 95\%, a latency of 0.5s, and an overhead of $1.8\times$ for \textit{every} node. (We compare with other schemes in \S\ref{sec:eval-comparison}.) %

We need simulations to evaluate \name in larger topologies. To verify the accuracy of the simulator, we run it with the same topology as the testbed, and compare the results with the aforementioned real-world experiments. \autoref{fig:sim-real-metrics} shows that for \textit{all} datapoints, results from simulations are within 1.5\% of that from the real-world testbed. %
To further demonstrate the accuracy of the simulator, \autoref{fig:sim-real-edges} shows a comparison of the rates that each node sends codewords to peers in the simulation and on the testbed. These rates are determined by the control algorithm (\S\ref{sec:design-multi-peers}), and reflect the operating state of each individual node in detail. For almost all pairs of peers (one sending codewords and the other receiving), the codeword rate in the simulation is within 3\% of that in the real-world experiment. These comparison show that the simulator provides reliable results by accurately modelling the operation of each individual node. From now on, we move to the simulator to explore larger topologies. We will come back to the real-world implementation in \S\ref{sec:eval-benchmark} to evaluate its computational efficiency.

\subsection{Performance comparison}
\label{sec:eval-comparison}

In this experiment, we compare the latency and the overhead of \name with other schemes. We use the 250-city topology. Each node generates transactions at 10 tps, resulting in a total throughput of 2500 tps.

\begin{figure}
\centering
\includegraphics[width=\columnwidth]{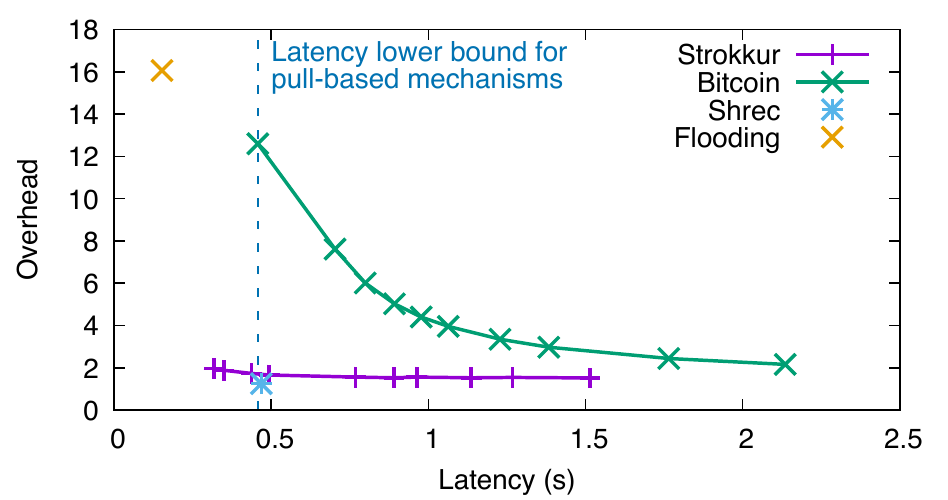}
\caption{Latency and overhead when all node are honest. Curves for \name and Bitcoin show possible trade-offs. For \name, we adjust the coding window size $k$ and the decoding timeout $\tau$. For Bitcoin, we adjust the max random jitter that nodes inject before forwarding the hash of a new transaction to peers. Dashed line shows the lowest latency that pull-based mechanisms (Bitcoin, Shrec) can achieve.}
\label{fig:main-comparison}
\end{figure}

\smallskip\noindent\textbf{Schemes compared.} We compare \name with three schemes: Bitcoin, Shrec, and transaction flooding. Transaction flooding is a simple scheme where nodes forward each new transaction to all peers. Bitcoin and Shrec are based on hash flooding, where nodes forward the hashes of newly-received transaction to peers, and explicitly request for unknown transactions from peers. Specifically, Bitcoin nodes inject a random jitter before forwarding the hash of a new transaction. Shrec, the state of the art scheme, enforces that each node only sends one request for each unknown transaction (hash). We describe these schemes in more detail in~\S\ref{sec:related-works}.

\smallskip\noindent\textbf{Transaction flooding.} Transaction flooding is a simple scheme where nodes forward transactions it receives for the first time to all neighbors. It achieves the lowest-possible latency among all schemes, because every transaction propagates as in a breadth-first search, and reaches each node along the shortest path possible. However, it incurs a high overhead of 16.0 because it makes no effort to avoid sending duplicated transactions. 
In comparison, \name can achieve $2\times$ latency while consuming $7\times$ less bandwidth, as shown in \autoref{fig:main-comparison}.

\smallskip\noindent\textbf{Bitcoin.} Both \name and Bitcoin provide a trade off between overhead and latency. For \name, we adjust the max codeword degree of the LT code between 10 and 100. A larger max degree reduces the overhead of the LT code \cite{robust-soliton-small-k}, but incurs a higher coding delay because transactions have to wait longer in the coding window before being selected into a codeword. For Bitcoin, we adjust the max random jitter before a node forwards the hash of a transaction to its neighbors. A larger max jitter spreads the hashes of a transaction across a wider timespan, so that a node is less likely to send duplicated requests for the same transaction, but it increases the latency. As \autoref{fig:main-comparison} shows, \name consumes $2\times$ less bandwidth than Bitcoin when both schemes achieve a latency of 1.5 seconds. As we push for a lower latency, the gap of the overhead between Bitcoin and \name increases, up to $7.6\times$ at a latency of 0.47 seconds.

\smallskip\noindent\textbf{Shrec.} Shrec nodes track the set of in-flight requests, and avoid sending duplicated requests for the same transaction. This technique works well when all nodes are honest and follow the protocol. Shrec achieves a latency of 0.47 seconds and an overhead of 1.25 in this scenario, using 27\% less bandwidth than \name. However, its performance relies on correct behavior of all nodes while blockchains networks with adversarial nodes. A malicious or malfunctioning node may not respond to transaction requests preventing them from concluding. %

\begin{figure}
\centering
\includegraphics[width=\columnwidth]{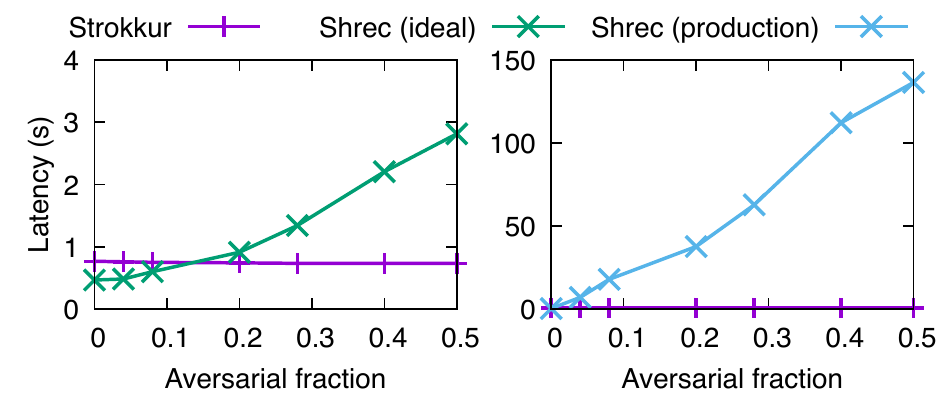}
\caption{Latency of \name and Shrec when a variable fraction of nodes are adversarial. Shrec (ideal) uses parameters tuned for the particular topology in this experiment. Such tuning is not practical for a real-world deployment. Shrec (production) uses the same parameters~\cite{conflux-parameters} as in the Conflux~\cite{conflux} blockchain.} %
\label{fig:robustness}
\end{figure}

Shrec does not specify a solution, and applications rely on heuristics to mitigate this issue. In its real-world deployment in the Conflux~\cite{conflux} blockchain, there is a timeout for transaction requests \cite{conflux-parameters}. If a request does not finish within the timeout, the requesting node retries with another neighbor. This timeout requires careful tuning. If set too high, a failed request can significantly delay a transaction. If set too low, a node may spuriously retry a request and receive duplicated responses, increasing the overhead. In \autoref{fig:robustness}, we compare \name and Shrec when a fraction of nodes are adversarial. Adversarial nodes do not send any codewords in \name, and do not respond to any requests in Shrec. We try two timeout settings for Shrec: a value fine-tuned for our topology, and the value used in production~\cite{conflux-parameters}. For the fine-tuned setting, we use the smallest timeout that ensures no node receives spurious transactions. This is an ideal setting that minimizes the latency. As the fraction of adversarial nodes increases, Shrec experiences higher latency, up to 3 seconds when 50\% of nodes are adversarial. We point out that such fine tuning is only possible in our experiments where the delay on each link is fixed, and the adversary does not try to attack the tuning process itself. In a real-world deployment, tuning is less practical. In practice, Conflux sets the timeout to 30 seconds~\cite{conflux-parameters}. Under this setting, the latency of Shrec increases to 7 seconds at only 4\% of adversarial nodes, and goes up to 136 seconds at 50\% adversarial nodes. In comparison, the latency of \name stays stable at 0.76 seconds across all scenarios.

\smallskip\noindent\textbf{Pull vs. push-based mechanisms.} At a high level, we evaluated two types of schemes in this subsection: pull-based, and push-based. In a pull-based scheme such as Bitcoin and Shrec, nodes first receive the hashes of transactions, and need to explicitly request for full transactions. It takes at least three single trips for a transaction to travel across a link. The vertical dashed line in \autoref{fig:main-comparison} illustrates this lower bound for the latency for such pull-based schemes. Bitcoin achieves it when the max jitter is set to 0, but incurs a high overhead of 12.59. Shrec achieves it only when all nodes are honest. In comparison, push-based schemes such as transaction flooding and \name achieve a lower latency, because nodes receive transaction data without sending any request.

\subsection{Censorship resilience}
\label{sec:eval-censorship}

\begin{figure}
\centering
\includegraphics[width=\columnwidth]{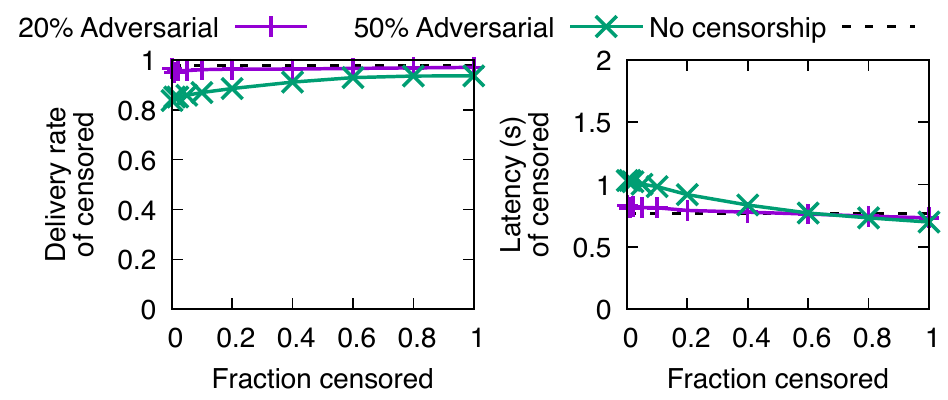}
\caption{Delivery rate and latency of censored transactions, when the adversary controls a fraction of nodes and censors a variable fraction of transactions.}
\label{fig:distributed-censorship}
\end{figure}

\begin{figure}
\centering
\includegraphics[width=\columnwidth]{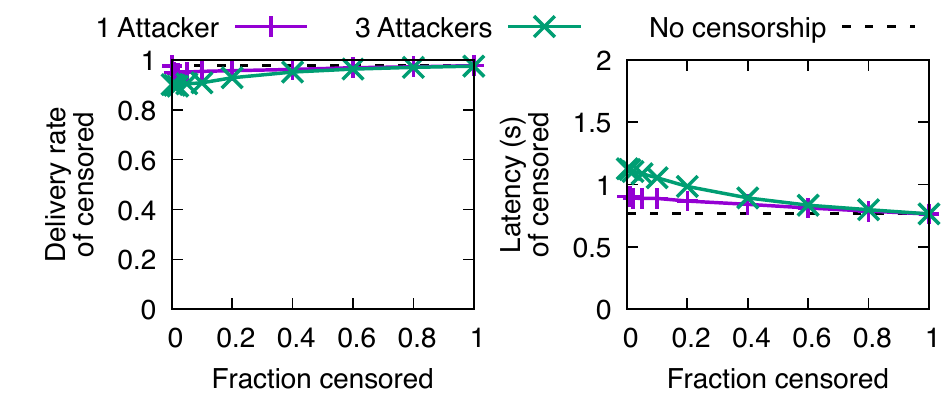}
\caption{Delivery rate and latency of censored transactions, when the adversary sets up attacker nodes with zero-delay connections to each of the honest nodes, and censors a variable fraction of transactions.}
\label{fig:central-censorship}
\end{figure}

In this experiment, we consider adversarial censoring of transactions. The topology and the workload is the same as the previous experiment. However, an adversary controls some nodes in the network, and attempts to prevent honest nodes from receiving certain transactions by not including them in the codewords they send. To determine how resilient \name is against such attacks, we measure the latency and the delivery rate of the censored transactions, i.e. the probability that an honest node can receive a censored transaction.  %
We consider two types of adversaries, positioning their adversarial nodes differently relative to the honest nodes. 

A distributed adversary controls a random subset of nodes in the network. 
\autoref{fig:distributed-censorship} shows the results of censorship for different portions of adversarial nodes in the network. When the adversary controls 20\% of the nodes, censored transactions have a delivery rate of 95\% and a latency of 0.83 second. Even a powerful adversary that controls 50\% of the nodes can only reduce the delivery rate to 83.7\%. Furthermore, the effectiveness of the attack depends on how many transactions the adversary tries to censor. To achieve the effectiveness we just described, the adversary cannot censor more than 0.1\% of the transactions. As this fraction increases, the effectiveness goes down. This is because the codeword rate controller on each honest node will (independently) notice that adversarial codewords have a lower loss rate (because they do not contain as many novel transactions---some of them are censored), and thus decrease their sending rates. Meanwhile, it will detect that the loss rate of honest codewords are higher (because they contain censored transactions), and increase the sending rates accordingly. For example, a 20\% adversary censoring 20\% of transactions succeeds with a probability of only 1.3\%, and increases the latency by only 0.02 seconds.

Next, we consider an even more powerful type of adversary, that sets up attacker nodes that connect to \textit{all} honest nodes with \textit{zero} latency. Such an adversary does not exist in the real world unless all honest nodes are co-located, but it serves as a worst-case scenario for evaluation. \autoref{fig:central-censorship} shows the delivery rate under such an attack. When there is one attacker node that connects to all honest nodes, censored transactions have a delivery rate of 95\% and a latency of 0.90 seconds. An adversary controlling three attacker nodes can reduce the delivery rate to 90\%. Similar to the previous type of adversary, this attack becomes less effective as the adversary tries to censor more transactions. In conclusion, both types of adversary cannot significantly affect the delivery rate and the latency of censored transactions, proving that \name is resilient to censorship.

\subsection{Scalability}
\label{sec:eval-scalability}

\begin{figure}
\centering
\includegraphics[width=\columnwidth]{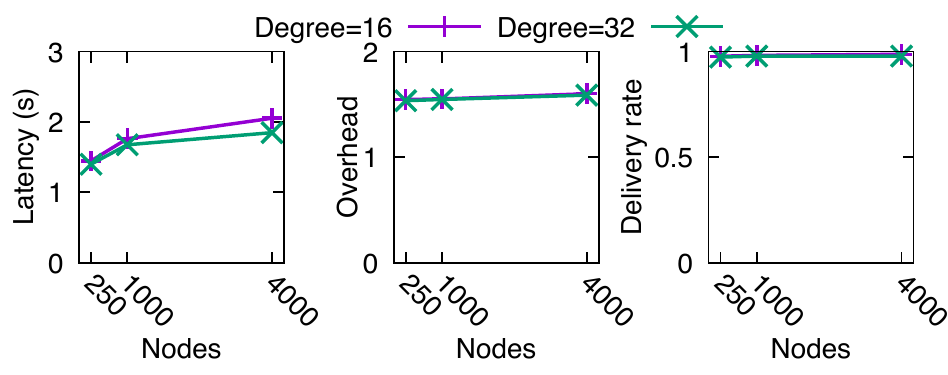}
\caption{Latency, overhead, and delivery rate in synthetic random regular topologies of degrees 16 and 32, and up to 4000 nodes.}
\label{fig:scalability}
\end{figure}

To demonstrate that \name retains its performance and efficiency in larger networks, we generate random regular graphs of 250, 1000, 4000 nodes with degrees 16 and 32. We sample the propagation delay (in milliseconds) of each link from a log-normal distribution with parameters $\mu=\ln(70)$ and $\sigma=0.5$. This distribution has a median of 70 and a standard deviation of 42.27, closely matching the median latency of 69.98 ms and the standard deviation of 45.56 ms of the 250-city data set that we use for the other experiments.
To ensure the simulations finish in reasonable time, we reduce the total transaction generation rate to 400 tps, and set the codeword timeout to 1 second. To show \name scales gracefully without any tuning, we fix all system parameters as we change topologies.

\autoref{fig:scalability} shows the results. As we increase the number of nodes from 250 to 4000, the latency goes up by 0.5s, because transactions have to travel across more links, experiencing more propagation delay and coding delay. The increased latency implies that nodes have to wait longer to receive enough transactions and decode a codeword. Meanwhile, recall that we do not attempt to tune the codeword timeout in this experiment, so more codewords fail to meet the decoding timeout, and cause the peers to increase the codeword rate. As a result, the overhead stays stable at 1.54 with 250 and 1000 nodes, then slightly increases to 1.58 with 4000 nodes. Increasing the codeword timeout to 2 seconds brings the overhead down to 1.55, but we do not think such fine tuning is worthwhile in most use cases. Finally, the transaction delivery rate stays stable at 98\% across all topologies.

Increasing the degree of the topology has long been proposed in the literature~\cite{hijack-bitcoin, erlay} to mitigate the eclipse attack~\cite{bitcoin-eclipse}, where the adversary controls all neighbors of a node to cut it from the network. On the other hand, it is known to cause higher overhead for many schemes since there are more potential neighbors to send duplicated transactions to a node. \name does not suffer from this issue. Results show that increasing the degree from 16 to 32 does not affect the overhead or the delivery rate, and reduces the latency because of the smaller network diameter. This is because the codeword rate controller ensures that nodes send codewords to each peer at the lowest rate that achieves the target loss rate, so that the system can always achieve the overhead of the underlying windowed LT code.

\subsection{Convergence}
\label{sec:eval-converge}

\begin{figure}
\centering
\includegraphics[width=\columnwidth]{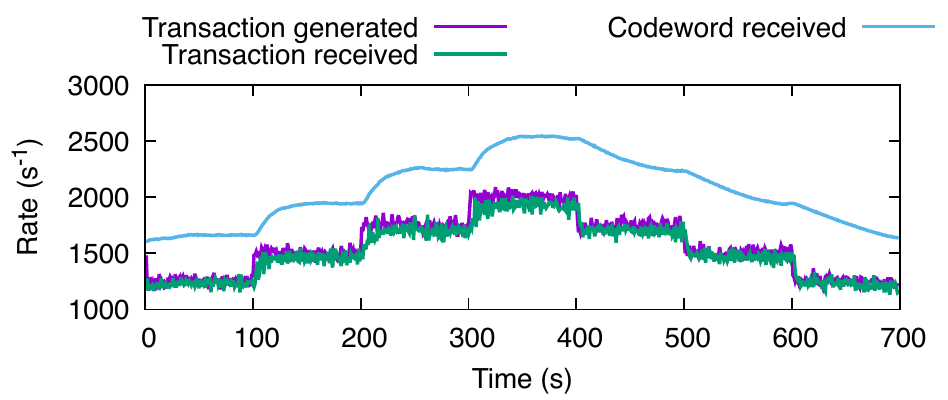}
\caption{Time series of transaction and codeword reception at a node, and transaction generation of the entire network.}
\label{fig:convergence}
\end{figure}

In this experiment, we show that the control algorithm in \name quickly converges when the load changes. We start the system with a transaction throughput of 1250 tps, increase the throughput to 2000 tps in three equal steps, and then reduce the throughput to 1250 tps in three steps. \autoref{fig:convergence} shows the trace of the experiment. When the throughput suddenly increases, more codewords begin to time out because there are not enough of them to accommodate the higher throughput. The timeout events trigger the controller to quickly ramp up the codeword sending rate, and the system converges within 10 seconds. When the throughput decreases, there are more codewords than needed for successful decoding. The controller sees fewer codeword timeouts than targeted, and reduces the codeword sending rate. Ramping down is slower than ramping up, because the controller decreases the codeword rate less aggressively than it does increase to avoid a catastrophic overshoot. Still, it takes only 100s for the network of 250 nodes to converge. Over the entire experiment, the system achieves a delivery rate of 97\%, an overhead of 1.37, and a latency of 0.75 seconds, similar to the performance when the load stays constant.

\subsection{Computational cost}
\label{sec:eval-benchmark}

Finally, we benchmark our implementation to show that encoding and decoding are computationally efficient, and can support high transaction throughputs. We use a 2018-model MacBook Pro with an Intel Core i7-8559U CPU and 16 GB of RAM. We set the coding window size $k$ to 50, and use the same settings for the other parameters as in previous experiments. We limit the node to only use one CPU core in the benchmarks.

\smallskip\noindent\textbf{Encoding.} We create and insert 50 random transactions into the coding window of a node, and measure the time for it to generate 800,000 codewords. The node can generate codewords at a throughput of 786.16 Mbps using only one CPU core.

\smallskip\noindent\textbf{Decoding.} We generate 800,000 random transactions, and encode them using the windowed LT code to obtain a stream of 1,080,000 codewords (1.35 codewords per transaction on average). We then feed the steam of codewords into a node, and measure the time for it to decode the codewords. The node can decode codewords at 895.2 Mbps, recovering transactions at 663.28 Mbps or 647,668 tps.%

These results show that \name is computationally efficient, and can support a throughput on the order of $10^{5}$ tps using only one CPU core. In comparison, future blockchains are likely to experience other bottlenecks such as verifying cryptographic signatures over transactions and executing transactions before \name's computational cost becomes the bottleneck. For example, a deployed implementation of the Ethereum transaction logic is computationally-bounded at 10,000 tps when executing simple payments~\cite{prism-evm}; a highly-optimized implementation of the Bitcoin transaction logic is IO-bounded at 80,000 tps~\cite{prism-system}. As we have demonstrated, \name can support such a throughput using less than one CPU core on a commodity laptop.

%% file: conclusion.tex
\section{Conclusion}
We presented \name, a new approach for broadcasting transactions in a blockchain's peer to peer network based on coding. 
We realize coding in an adversarial network by extending the LT code to support a continuous stream of transactions. We designed an effective control algorithm that allows nodes to exchange codewords at the optimal rate that minimizes the communication cost and maintains a low latency. 
Through evaluation on a prototype implementation and simulations, we showed that \name
quickly adapts codeword transmission rates across peers to the transaction rates to minimize loss, resists adversarial nodes trying to degrade performance or censor transactions, scales well to large networks, and has low computational costs. These properties make \name suitable for high performance blockchain networks. The design is useful as a generic scheme to broadcast short messages in an adversarial network with low overhead and low latency.